%% file: mnras_template.tex
\documentclass[fleqn,usenatbib]{mnras}

\usepackage{newtxtext,newtxmath}

\usepackage[T1]{fontenc}
\usepackage{ae,aecompl}

\usepackage{graphicx}	
\usepackage{amsmath}	
\usepackage{amssymb}	

\input{properties.tex}

\newcommand{\rev}{\textcolor{black}}

\newcommand{\longobjname}{ULAS J224940.13-011236.9}
\newcommand{\objname}{ULAS J2249-0112}
\newcommand{\startdate}{2017 May 07}
\newcommand{\stopdate}{2017 Dec 14}
\newcommand{\flarenight}{2017 Aug 13}
\newcommand{\nightnumber}{146}
\newcommand{\Gaia}{\textit{Gaia}}
\newcommand{\kepler}{\textit{Kepler}}

\newcommand{\Kepler}{\textit{Kepler}}

\newcommand{\TESS}{\textit{TESS}}

\newcommand{\Ktwo}{\textit{K2}}

\newcommand{\energyvalue}{\rev{$3.4^{+0.9}_{-0.7}\times10^{33}$erg}}
\newcommand{\carringtonenergy}{$10^{32}$ erg}
\newcommand{\visibleduration}{9.5 minutes}
\newcommand{\efoldtime}{2.1 minutes}
\newcommand{\thalfvalue}{1.5 minutes}
\newcommand{\thalf}{$t_{1/2}$}
\newcommand{\halpha}{H$\alpha$}
\newcommand{\energytime}{\rev{1.8}}
\newcommand{\flareemittingarea}{\rev{$5.6\times10^{19}$cm$^2$}}
\newcommand{\flareemittingpercent}{\rev{$37$}} 

\title[Detection of an L2.5 dwarf superflare]{Detection of a giant white-light flare on an L2.5 dwarf with the Next Generation Transit Survey}

\author[J. A. G. Jackman]{
James A. G. Jackman,$^{1,2}$\thanks{E-mail: J.Jackman@warwick.ac.uk}
Peter J. Wheatley,$^{1,2}$\thanks{E-mail: P.J.Wheatley@warwick.ac.uk}
Daniel Bayliss,$^{1,2}$\newauthor
Matthew R. Burleigh,$^3$ 
Sarah L. Casewell,$^3$
Philipp Eigm\"uller,$^{4,5}$\newauthor
Mike R. Goad,$^3$
Don Pollacco, $^{1,2}$
Liam Raynard,$^3$
Christopher A. Watson,$^{6}$\newauthor
Richard G. West,$^{1,2}$
\\ \\
$^{1}$Dept. of Physics, University of Warwick, Gibbet Hill Road, Coventry CV4 7AL, UK \\
$^{2}$Centre for Exoplanets and Habitability, University of Warwick, Gibbet Hill Road, Coventry CV4 7AL, UK\\
$^{3}$Department of Physics and Astronomy, University of Leicester, University Road, Leicester, LE1 7RH\\
$^{4}$ Institute of Planetary Research, German Aerospace Center, Rutherfordstrasse 2, 12489 Berlin, Germany\\ 
$^{5}$ Center for Astronomy and Astrophysics, TU Berlin, Hardenbergstr. 36, D-10623 Berlin, Germany\\
$^{6}$ Astrophysics Research Centre, School of Mathematics and Physics, Queen's University Belfast, Belfast, BT7 1NN, UK
}

\date{Accepted XXX. Received YYY; in original form ZZZ}

\pubyear{2019}

\begin{document}
\label{firstpage}
\pagerange{\pageref{firstpage}--\pageref{lastpage}}
\maketitle

\begin{abstract}
We present the detection of a $\Delta V\sim$ -10 flare from the ultracool L2.5 dwarf \longobjname\ with the Next Generation Transit Survey (NGTS). The flare was detected in a targeted search of late-type stars in NGTS full-frame images and represents one of the largest flares ever observed from an ultracool dwarf. 
This flare also extends the detection of white-light flares to stars with temperatures below 2000\,K. 
We calculate the energy of the flare to be \energyvalue, making it an order of magnitude more energetic than the Carrington event on the Sun. Our data show how the high-cadence NGTS full-frame images can be used to probe white-light flaring behaviour in the latest spectral types.
\end{abstract}




\section{Introduction}
Previous studies of L dwarfs \citep[\teff 1300-2300\,K][]{Stephens09} have shown them to be variable in a number of ways. Examples of this include periodic modulation due to clouds \citep[e.g][]{Gizis_clouds}, radio emission due to aurora in late L dwarfs \citep[e.g.][]{Kao18} and the presence of white-light flares \citep[e.g.][]{Gizis13}. white-light flares occur through reconnection events in the stellar magnetic field, which result in heating of the lower chromosphere/ upper photosphere \citep[e.g.][]{Benz10}, resulting in white-light emission. While seen regularly on GKM stars, observations of white-light flares on L dwarfs remain rare, with only a handful of stars showing them to date \citep[e.g.][]{Paudel18}. However, those observed have included some of the largest amplitude flares ever recorded, reaching up to $\Delta V\approx-11$ \citep[][]{Schmidt16}. This shows that white-light flaring activity persists into the L spectral type, despite previous studies of L dwarfs showing their chromospheres and magnetic activity to be diminished compared to those of late M dwarfs \citep[e.g][]{Schmidt15}.

Due to the rarity of these large amplitude flares, long duration observations are required when targeting L dwarfs specifically. This became possible with the \kepler\ \citep[][]{Borucki10} and \Ktwo\ \citep[][]{Howell14} missions, which allowed continuous high precision monitoring of chosen objects \citep[][]{Gizis13,Gizis17,Paudel18}. Through this, large amplitude flares ($\Delta V \approx -8$, $\Delta K_{p}=-5.4$) were detected on an L0 and L1 dwarf \citep[][]{Gizis17,Paudel18}, along with smaller flares on two early L dwarfs \citep[][]{Gizis13,Paudel18}. \rev{Spectroscopy of an L5 dwarf by \citet{Liebert03} 
showed an \halpha\ flare, 
although this event 
lacked the continuum enhancement associated with white light flares. Observations of an L5 dwarf \citep[][]{Gizis17b} with \Ktwo\ showed no detectable white light flares, suggesting such events are suppressed at late spectral types.} For targeted observations it is typically required that the star be visible in quiescence. For optical surveys which are not sensitive to late spectral types, this requirement limits observations to a small number of nearby early L dwarfs. 

Wide-field surveys (e.g. ASAS-SN), which obtain full frame images of large areas of sky per night allow studies to push to fainter stars. In these surveys, full frame images can be used to search for very faint stars that only become detectable when flaring. This led to the detection of a $\Delta V \approx -11$ flare from the L0 dwarf ASASSN-16ae \citep[][]{Schmidt16}, previously the only L dwarf flare detected from the ground \rev{with optical photometry}. 
While these surveys can detect large flares, the low cadence lightcurves limit the study of shape and substructure, and hence also the accuracy 
of flare energy calculations \citep[e.g.][]{Schmidt18}. Consequently, in order to detect and characterise the largest flare events on the latest spectral types, high cadence wide-field observations with full-frame images are required.

In this letter we present the detection of a white-light superflare from the L2.5 dwarf \longobjname, which was observed at high cadence with the Next Generation Transit Survey (NGTS). This source has also previously been identified as \twomassname. 
We describe the detection of the flare and how we derived the properties of the flare and the host star. 

\section{Observations}

NGTS is a ground-based wide-field exoplanet survey, which obtains full-frame images from twelve independent telescopes at 12\,s cadence \citep[][]{Wheatley18}. The telescopes have apertures of 20\,cm and operate with a bandpass of 520-890nm. They have a total instantaneous field of view of 96 square degrees.

The data presented in this letter were collected with NGTS between \startdate\ and \stopdate, comprising \nightnumber\ nights of data. The detected flare occurred on the night of \flarenight. 

\begin{figure}
	\includegraphics[width=\columnwidth]{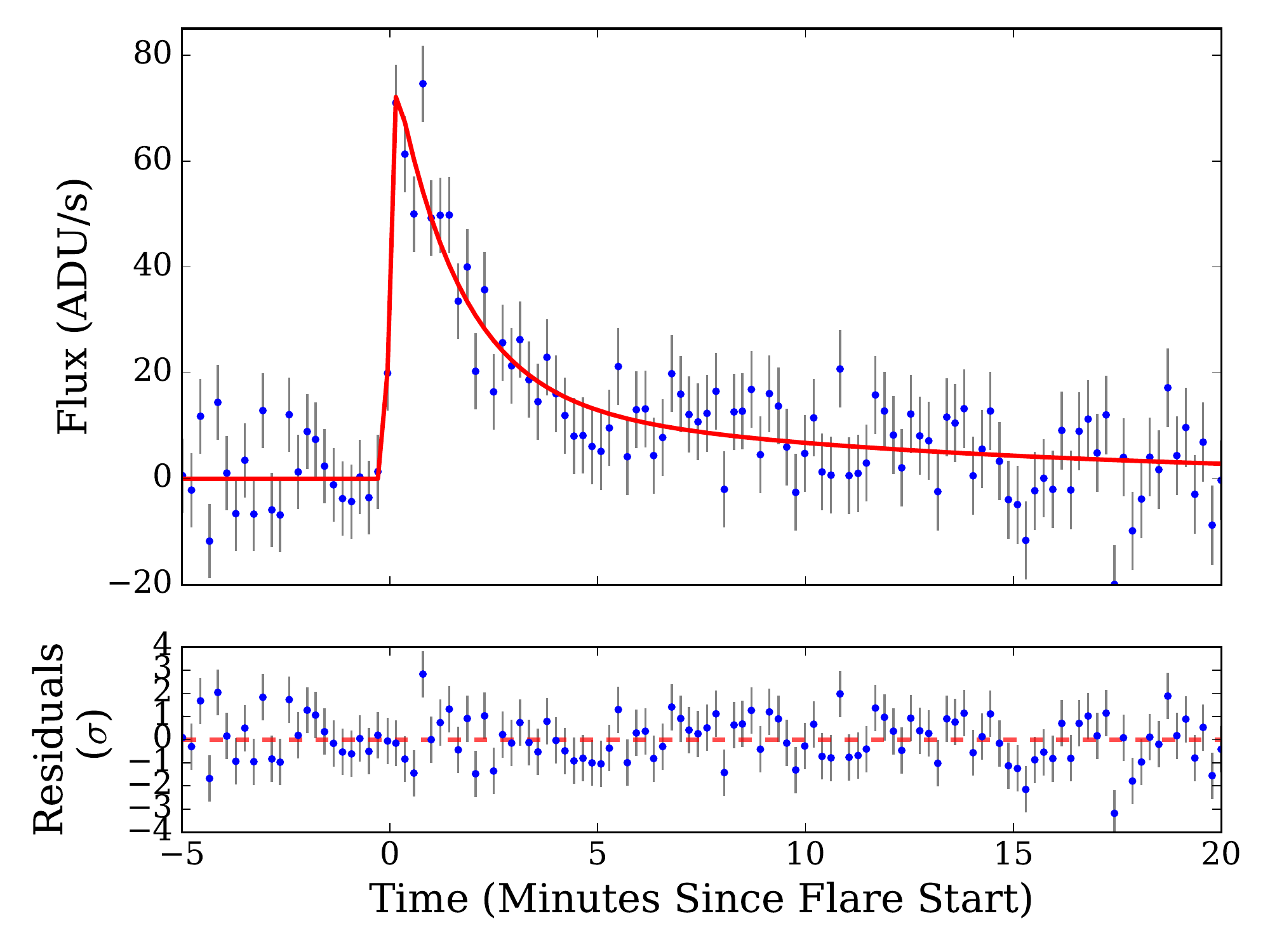}
    \caption{Top: NGTS lightcurve of the flare from \objname. Blue points are NGTS photometry and the best fitting model is overlaid in red. Note the outlier near the flare peak, something we discuss further in Sect.\,\ref{sec:structure}. Bottom: Residuals from the model fit.}
    \label{fig:Flare_fit}
\end{figure}

\subsection{Input Catalogue and Flare Detection}
In the normal mode of operations NGTS obtains lightcurves for all stars brighter than I = 16 \citep[][]{Wheatley18}. However, in each NGTS field there exist on average 2500 red stars fainter than this limit which can be identified using 2MASS \citep[][]{2MASS_2006} and \Gaia\ \citep[][]{Gaia18}. While too faint to observe as primary targets, by placing photometric apertures on their positions we can capture any variability, such as flares, that may result in them becoming bright enough to detect. 
To this end, for each NGTS field we compiled input catalogues of known red stars using positions from 2MASS and the colour cuts for ultracool dwarfs specified by \citet{Muirhead18}. The normal NGTS pipeline \citep[][]{Wheatley18} was then run on these positions. To search for flares we follow the procedure of \citet{Jackman18a} and look for consecutive outliers 6 median absolute deviations (MAD) above the median of a single night.

Using this method we detected a flare from the star \longobjname\ \citep[][]{Skrzypek16} (hereafter \objname),
shown in Fig.\,\ref{fig:Flare_fit}.  To confirm this star is the source of the flare we checked the centroiding of this source, along with the NGTS images before, during and after the flare. We can see no measurable centroid shift during the flare and do not identify any brightening from a nearby source or from a satellite passing through our aperture. Consequently, we are confident the detected flare comes from \objname.

\subsection{Stellar Properties}
\objname\ has previously been identified as an L2.5($\pm1$) dwarf by \citet{Skrzypek16}, as part of a sample of photometrically classified L and T dwarfs from SDSS \citep[][]{York2000}, UKIDSS \citep[][]{Lawrence07} and WISE \citep[][]{Wright10}. To confirm this and measure the effective temperature of \objname\ we fit the spectral energy distribution (SED) using the BT-Settl models \citep[][]{Allard12}. We fit to the catalogue photometry in Tab.\,\ref{tab:catalogue_photometry}, following a similar method to \citet{Gillen17}. We note here that this source is too faint to be detected by \Gaia. From our fitting we measure an effective temperature of \teff=\teffvalue, consistent with that expected for the L2.5 spectral type \citep[][]{Stephens09}. 

\objname\ was also observed by the Baryon Oscillation Spectroscopic Survey \citep[BOSS;][]{Dawson13}. BOSS obtains optical spectra with a resolution of R $\approx$ 2000 over a wavelength range 3600-10000\,\AA. 
\objname\ also sits in SDSS stripe 82, however due to its faint \textit{i} magnitude was not included in the analysis of BOSS ultracool dwarfs by \citet{Schmidt15}. We excluded wavelengths below 5400\,\AA\ in our analysis due to the lack of emission in the blue from \objname. 

\begin{figure}
	\includegraphics[width=\columnwidth]{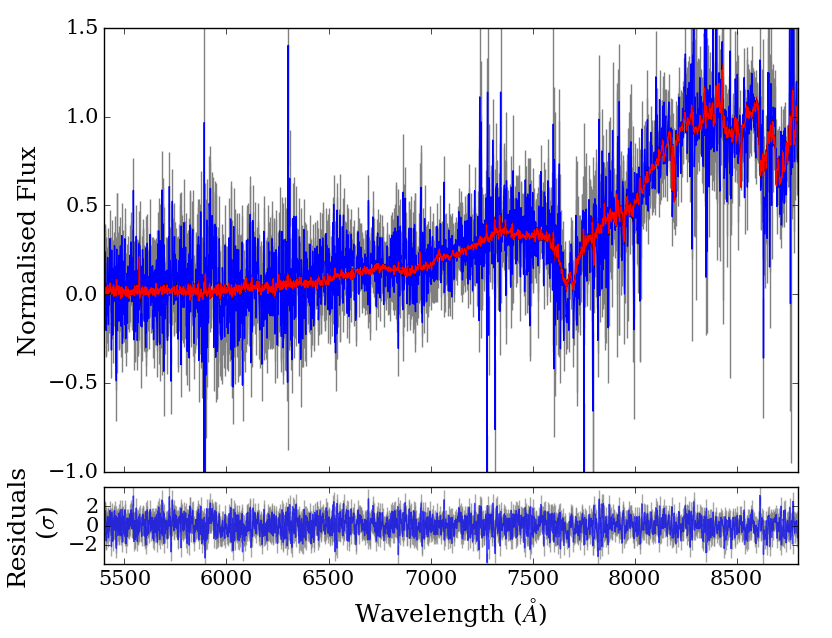}
    \caption{Top: BOSS spectrum of \objname\ in blue, with errors in grey. The spectrum has been normalised at 8350--8450\,\AA\ with the best-fitting L-dwarf template overlaid in red. Bottom: Residuals from the fit.}
    \label{fig:BOSS_spectrum}
\end{figure}

Fitting this spectrum using the empirical L dwarf templates of \citet{Schmidt15} gives a best fitting spectral type of L2.6$\pm$0.1, again confirming 
our spectral type determination and the low temperature. 
The BOSS spectrum with the best fitting template is shown in Fig.\,\ref{fig:BOSS_spectrum}. Due to the low S/N ratio (1.2 around H$\alpha$) of the spectrum we are not able to clearly identify the presence of H$\alpha$ emission in quiescence.

\subsubsection{Kinematics}
In order to estimate the distance to \objname\ we initially calculate the absolute 2MASS and WISE magnitudes using the relations of \citet{Dupuy12}, derived from ground-based infrared astrometry, using the L2.5 spectral type. When calculating the distances using these individual absolute magnitudes, we found all calculated distances to be consistent. We thus took a weighted mean of the individual photometric distances, which we calculate as \distancevalue. 
For the proper motion we have used the measurements of \citet{Bramich08}, which are given in Tab.\,\ref{tab:catalogue_photometry}. These values were obtained using the repeat measurements of \objname\ as part of SDSS stripe 82. For the radial velocity we have used the value provided with the BOSS spectrum, $v_{rad}=-14\pm10$\,km\,s$^{-1}$. By combining these values together we calculate the three-dimensional velocities (\textit{U},\textit{V},\textit{W})=(\uvel, \vvel, \wvel)\,km\,s$^{-1}$. These values appear to place \objname\ within the thin disc \citep[][]{Schmidt10,Burgasser15}. 

\begin{table}
 \begin{tabular}{lcc}
  \hline
  Property & Value & Reference\\
  \hline
  R.A (J2000, Deg)  & \ravalue\ & 1 \tabularnewline
  Dec (J2000, Deg) & \decvalue\ & 1 \tabularnewline
  $i'$ & \imag\ & 2 \tabularnewline
  $z'$ & \zmag\ & 2 \tabularnewline
  $J$ & \jmag\ & 1 \tabularnewline
  $H$ & \hmag\ & 1 \tabularnewline
  $K_{s}$ & \kmag\ & 1 \tabularnewline
  $W1$ & \wonemag & 3 \tabularnewline
  $W2$ & \wtwomag\ & 3 \tabularnewline
  $\mu_{R.A.}$ (mas yr$^{-1}$) & \pmra & 2 \tabularnewline
  $\mu_{Dec}$  (mas yr$^{-1}$) & \pmdec & 2 \tabularnewline
  \teff\ (K) & \teffvaluenoK & This work \tabularnewline
  Distance (pc) & \distancevaluenopc & This work \tabularnewline
  Radius ($R_{\odot}$) & \radiusvalue & This work \tabularnewline
  \hline
 \end{tabular}
 \caption{Properties of \longobjname. References: 1. \citet{2MASS_2006}, 2. \citet{Bramich08}, 3. \citet{ALLWISE2014}.
  \label{tab:catalogue_photometry}}
\end{table}

\section{Results}

\subsection{Amplitude and Duration} \label{sec:amp_dur}
To determine the amplitude and duration of the flare, and to search for signs of substructure, we fit the NGTS data with the solar-flare inspired model from \citet{Jackman18a}. This model 
incorporates both a Gaussian heating term and a double exponential for the flare cooling. We fit the model using an MCMC process with 100 walkers for 20000 steps. We disregard the first 1000 steps as a burn-in. The result of this fit is shown in Fig.\,\ref{fig:Flare_fit}. From looking at the residuals to the fit we can identify a single outlier around the flare peak which may be signs of partially resolved flare substructure, something we discuss further in Sect.\,\ref{sec:structure}. 
We use the best fitting model to measure the duration for which the flare is at least 1$\sigma$ above the background, where $\sigma$ is the standard deviation of the lightcurve when the star is in quiescence.  We measure the visible duration as \visibleduration. We can use also use our fit to measure the e-folding timescale and the $t_{1/2}$ duration of the flare. These parameters are often used in flare analysis \citep[e.g.][]{Shibayama13,Davenport14} and provide separate measures of the duration. We measure these as 2.1 and 1.5 minutes respectively. 
The FWHM of the Gaussian heating term also gives a measure of the flare rise-time, something not typically measured in other surveys due to its short duration. We find this to be 12 seconds, indicating rapid heating.

We can also use our fitted model to measure the amplitude of the flare, using the peak count rate. 
In order to determine what magnitude this corresponds to (in order to calculate the fractional amplitude of the flare) we have crossmatched all NGTS sources in this field with SDSS to obtain a linear relation between the NGTS instrumental magnitude and SDSS $i'$ band. Using this relation we convert the NGTS instrumental flux to a proxy $i'$ band magnitude of 15.3, resulting in $\Delta i'\simeq -6$. For comparison with previous works on large L dwarf flares, we also estimate the magnitude in the V band. To do this we use the best fitting BT-Settl model and assume the flare is given by a $9000\pm500$K blackbody-like emitter \citep[e.g.][]{Hawley1992}. From this we estimate $\Delta V\sim -10$ and $\Delta U\sim -15$. This is consequently the second largest observed white-light flare from an L dwarf, following ASASSN-16ae \citep[$\Delta V$=-11][]{Schmidt16}, and larger than those observed with \Kepler\ \citep[][]{Gizis13,Gizis17,Paudel18}.

\subsection{Flare Energy} \label{sec:flare_energy}
To calculate the flare energy we have again assumed the flare is given by a $9000\pm500$\,K blackbody. \rev{We initially renormalise the blackbody to match the SDSS $i'$ band magnitude for each point in the calibrated flare amplitude lightcurve from Sect.,\ref{sec:amp_dur}.}
For each time step we integrate the corresponding \rev{renormalised} flare blackbody over all wavelengths and multiply by $4\pi d^{2}$, where $d=$ \distancevalue, to calculate the bolometric flare luminosity. 
Finally we integrate over the visible duration to calculate the total bolometric energy of the flare. This results in a bolometric flare energy of \energyvalue, which is over an order of magnitude greater than the energy of the Carrington event on the Sun \citep[\carringtonenergy][]{Carrington_1859,Carrington_Energy}. Another way of analysing this is to compare it to the total bolometric luminosity of the star in quiescence. Integrating over our best fitting SED results in a luminosity of $5.3\times10^{29}$erg s$^{-1}$, meaning the flare is equivalent in energy to \energytime\ hours of quiescent emission from \objname. We note that this energy is only for the visible duration of the flare and is thus a lower limit.

\section{Discussion}
We have detected a large amplitude white-light superflare on the \spectype\ star \objname. This is only the second L-dwarf flare to be detected from the ground and the sixth L-dwarf to have exhibited flaring activity. It is the coolest star ever found to exhibit a white-light flare. The flare had an amplitude of $\Delta V\sim$ -10 and an energy of \energyvalue, making it the second largest amplitude L dwarf flare, following that of ASASSN-16ae \citep{Schmidt16}. With a cadence of 12 seconds this is the best resolved detection of a giant white-light flare, allowing direct measurements of the amplitude and duration without relying on extrapolation. 
\subsection{Magnetic Activity of L dwarfs}
The detection of a white-light flare from an L2.5 dwarf makes this the coolest star to show such an event and shows strong chromospheric activity can persist to this spectral type. 
Studies of the H$\alpha$ emission from ultracool dwarfs using BOSS spectra \citep[][]{Schmidt15} has found the activity strength log$\big(L_{H\alpha}/L_{bol}\big)$ to decrease from approximately -3.8 at the M4 spectral type towards -5.7 at L3. This decrease in H$\alpha$ activity may imply the chromospheres of early L dwarfs are significantly cooler than their late-M dwarf counterparts, or cover a smaller fraction of the surface (lower filling factor) \citep[][]{Schmidt15}. Recent discoveries of mid-L dwarfs with H$\alpha$ emission have shown this decrease in chromospheric strength seems to continue with spectral type \citep[e.g.][]{Perez17} and can be linked to the 
decreasing ionisation in the L dwarf photosphere \citep[e.g.][]{Miles17}. 
Ultracool dwarfs are also known to exhibit auroral activity \citep[e.g.][]{Kao18}, which may account for observed H$\alpha$ emission in these systems. It is expected that the transition from predominantly chromospheric to auroral H$\alpha$ emission occurs during the L spectral type \citep[][]{Pineda17}. Many ultracool dwarfs which show activity such as radio emission and flaring also tend to be fast rotators, with rotation periods on the order of hours. However, neither the L1 dwarf \gizislongname\ \citep[][]{Gizis17} nor the L0 dwarf \paudellongname\ \citep[][]{Paudel18} showed any sign of rapid rotation when observed by \Ktwo, despite showing large amplitude white-light flares. Consequently, we do not attempt to predict whether \objname\ is a fast rotator. Regardless of this, studies of white-light flares such as from \objname\ can aid in understanding exactly how far into the L spectral type chromospheric activity persists.

\subsection{Flare Structure} \label{sec:structure}
In Sect.\,\ref{sec:amp_dur} we fitted the flare using a single continuous model, shown in Fig.\,\ref{fig:Flare_fit}.   
This model incorporates a Gaussian heating pulse with two exponential decays, for thermal and non-thermal cooling respectively \citep{Jackman18a}. 
From our fit we can also identify the presence of a possible second peak. This is from an outlier lying 2.8\,$\sigma$ from the fitted model and occurring approximately 40 seconds after the fitted peak.  
Multiple peaks in flares have previously been attributed to flares occurring on different parts of the star and previous studies have found that higher energy flares are more likely to be complex \citep[e.g.][]{Hawley14}. One possibility for multiple peaks is sympathetic flaring, where one flare triggers flaring in different active regions \citep[e.g.][]{Moon02}. 
By assuming the flare is given by a 9000 K blackbody we can estimate the maximum area of emission. We estimate our flare has a maximum emitting area of \flareemittingarea, or \flareemittingpercent\ per cent of the visible hemisphere \citep[e.g.][]{Hawley04}. 
This value for the emitting area is similar to those for flares on M dwarfs \citep[e.g.][]{Osten10}, however covers a much greater fraction of the surface. For example, a flare of similar energy observed by \citet{Hawley04} from the M3.5 dwarf AD Leo (which had multiple peaks) had a covering fraction of $\approx$ 1 per cent. If we assume this emission area is related to the area of the flare footpoints \citep[as in][]{Osten10} and that only flares of similar energy would cause visible substructure, then multiple flares with large footpoints on the stellar surface in quick succession would be required. This requirement may inhibit the presence of multiple peaks in L dwarf flares, suggesting the observed outlier is perhaps not due to a separate event. If this is the case, then the lack of complexity would make it similar to the $\Delta K_p$=\,-5.4 flare detected by \citet{Paudel18} on the L0 dwarf 2MASS J12321827-0951502. This flare showed no obvious substructure other than an extended decay when observed in the \Ktwo\ 1 minute short cadence mode.

In Sect.\,\ref{sec:amp_dur} we measured the visible duration of \objname\ as \visibleduration. Along with this, we were able to measure the e-folding duration as \efoldtime, the \thalf\ duration as \thalfvalue\ and the FWHM of the Gaussian heating pulse as 12 seconds.  Comparing the \thalf\ duration to other L dwarf flares, we find it is smaller than the three other large amplitude L dwarf flares \citep[][]{Schmidt16,Gizis17,Paudel18} and is more comparable to the smaller flares observed from the L1 dwarf WISEP J190648.47+401106.8 \citep[e.g.][]{Gizis13}. While this implies that this is the most rapid large amplitude flare observed from an L dwarf, we note that all large amplitude L dwarf flares have had \thalf\ values below 10 minutes. As such, other observations may have been hindered by the resolution of observations (as noted by \citet{Paudel18}) which would smear out flares in time \citep[e.g][]{Yang18}. \rev{To estimate this effect we bin our data to cadences of 1 and 2 minutes, to simulate the \kepler\ and \TESS\ short cadence modes. We measure \thalf\ timescales of 2 and 3.25 minutes respectively, showing that  \thalf\ measurements can be significantly affected by the longer cadence of these space telescopes.} The short duration of our flare thus highlights the requirement of high cadence observations in order to detect and characterise these events. This behaviour would not be visible in the \TESS\ 30 minute cadence full frame images \citep[][]{Ricker15}, nor in all-sky surveys which monitor large areas of sky each night. This shows how NGTS is in an ideal position to probe flare behaviour on the latest spectral types.


\section{Conclusions}
In this work we have presented the detection of a giant white-light flare from the \spectype\ star \longobjname. The flare was detected from a dedicated search for stellar flares on low mass stars in the NGTS full frame images and has an amplitude of $\Delta i'\approx$ -6, or $\Delta V\sim$ -10. This makes it the second largest flare to be detected from an L dwarf and is the second to be detected from the ground. With a spectral type of L2.5 we believe \objname\ is the coolest 
star to show a white-light flare to date. This flare detection also highlights the value of the NGTS high-cadence full-frame images in studying the largest stellar flares from the coolest stars.

\section*{Acknowledgements}
We are grateful to the referee, John Gizis, for their helpful suggestions. This research is based on data collected under the NGTS project at the ESO La Silla Paranal Observatory. The NGTS facility is funded by a consortium of institutes consisting of 
the University of Warwick,
the University of Leicester,
Queen's University Belfast,
the University of Geneva,
the Deutsches Zentrum f\" ur Luft- und Raumfahrt e.V. (DLR; under the `Gro\ss investition GI-NGTS'),
the University of Cambridge, together with the UK Science and Technology Facilities Council (STFC; project reference ST/M001962/1). 
JAGJ is supported by STFC PhD studentship 1763096.
PJW is supported by STFC consolidated grant ST/P000495/1.
PE acknowledge the support of the DFG priority program SPP 1992 "Exploring the Diversity of Extrasolar Planets" (RA714/13-1).
This publication makes use of data products from the Two Micron All Sky Survey, which is a joint project of the University of Massachusetts and the Infrared Processing and Analysis Center/California Institute of Technology, funded by the National Aeronautics and Space Administration and the National Science Foundation.
This publication makes use of data products from the Wide-field Infrared Survey Explorer, which is a joint project of the University of California, Los Angeles, and the Jet Propulsion Laboratory/California Institute of Technology, funded by the National Aeronautics and Space Administration.




\bibliographystyle{mnras}
\bibliography{references} 





\bsp	
\label{lastpage}
\end{document}

%% file: properties.tex
\newcommand{\teff}{$T_{\rm eff}$}
\newcommand{\teffvalue}{$1930\pm 100$\,K}
\newcommand{\teffvaluenoK}{$1930\pm 100$}

\newcommand{\ravalue}{$342.417115$}
\newcommand{\decvalue}{$-1.210352$}
\newcommand{\pmra}{$87.4\pm4.2$}
\newcommand{\pmdec}{$22.6\pm4.2$}
\newcommand{\spectype}{L2.5}
\newcommand{\distancevalue}{$76^{+8}_{-7}$ pc}
\newcommand{\distancevaluenopc}{$76^{+8}_{-7}$}
\newcommand{\radiusvalue}{$0.10\pm0.02$}
\newcommand{\uvel}{$-33.5\pm4.2$}
\newcommand{\vvel}{$-10.3\pm6.1$}
\newcommand{\wvel}{$0.5\pm7.9$}
\newcommand{\twomassname}{2MASS J22494010-0112372}

\newcommand{\gizislongname}{SDSSp J005406.55-003101.8}
\newcommand{\paudellongname}{J12321827-0951502}
\newcommand{\jmag}{$16.832\pm0.138$}
\newcommand{\hmag}{$16.065\pm0.177$}
\newcommand{\kmag}{$15.610\pm0.215$}
\newcommand{\wonemag}{$15.091\pm0.037$}
\newcommand{\wtwomag}{$14.775\pm0.082$}
\newcommand{\imag}{$21.461\pm0.101$}
\newcommand{\zmag}{$19.627\pm0.102$}


%% file: mnras_template.bbl
\begin{thebibliography}{}
\makeatletter
\relax
\def\mn@urlcharsother{\let\do\@makeother \do\$\do\&\do\#\do\^\do\_\do\%\do\~}
\def\mn@doi{\begingroup\mn@urlcharsother \@ifnextchar [ {\mn@doi@}
  {\mn@doi@[]}}
\def\mn@doi@[#1]#2{\def\@tempa{#1}\ifx\@tempa\@empty \href
  {http://dx.doi.org/#2} {doi:#2}\else \href {http://dx.doi.org/#2} {#1}\fi
  \endgroup}
\def\mn@eprint#1#2{\mn@eprint@#1:#2::\@nil}
\def\mn@eprint@arXiv#1{\href {http://arxiv.org/abs/#1} {{\tt arXiv:#1}}}
\def\mn@eprint@dblp#1{\href {http://dblp.uni-trier.de/rec/bibtex/#1.xml}
  {dblp:#1}}
\def\mn@eprint@#1:#2:#3:#4\@nil{\def\@tempa {#1}\def\@tempb {#2}\def\@tempc
  {#3}\ifx \@tempc \@empty \let \@tempc \@tempb \let \@tempb \@tempa \fi \ifx
  \@tempb \@empty \def\@tempb {arXiv}\fi \@ifundefined
  {mn@eprint@\@tempb}{\@tempb:\@tempc}{\expandafter \expandafter \csname
  mn@eprint@\@tempb\endcsname \expandafter{\@tempc}}}

\bibitem[\protect\citeauthoryear{{Allard}, {Homeier}  \& {Freytag}}{{Allard}
  et~al.}{2012}]{Allard12}
{Allard} F.,  {Homeier} D.,   {Freytag} B.,  2012, \mn@doi [Philosophical
  Transactions of the Royal Society of London Series A]
  {10.1098/rsta.2011.0269}, \href
  {https://ui.adsabs.harvard.edu/#abs/2012RSPTA.370.2765A} {370, 2765}

\bibitem[\protect\citeauthoryear{{Benz} \& {G{\"u}del}}{{Benz} \&
  {G{\"u}del}}{2010}]{Benz10}
{Benz} A.~O.,  {G{\"u}del} M.,  2010, \mn@doi [Annual Review of Astronomy and
  Astrophysics] {10.1146/annurev-astro-082708-101757}, \href
  {https://ui.adsabs.harvard.edu/\#abs/2010ARA&A..48..241B} {48, 241}

\bibitem[\protect\citeauthoryear{{Borucki} et~al.,}{{Borucki}
  et~al.}{2010}]{Borucki10}
{Borucki} W.~J.,  et~al., 2010, \mn@doi [Science] {10.1126/science.1185402},
  \href {https://ui.adsabs.harvard.edu/\#abs/2010Sci...327..977B} {327, 977}

\bibitem[\protect\citeauthoryear{{Bramich} et~al.,}{{Bramich}
  et~al.}{2008}]{Bramich08}
{Bramich} D.~M.,  et~al., 2008, \mn@doi [\mnras]
  {10.1111/j.1365-2966.2008.13053.x}, \href
  {https://ui.adsabs.harvard.edu/\#abs/2008MNRAS.386..887B} {386, 887}

\bibitem[\protect\citeauthoryear{{Burgasser} et~al.,}{{Burgasser}
  et~al.}{2015}]{Burgasser15}
{Burgasser} A.~J.,  et~al., 2015, \mn@doi [The Astrophysical Journal Supplement
  Series] {10.1088/0067-0049/220/1/18}, \href
  {https://ui.adsabs.harvard.edu/\#abs/2015ApJS..220...18B} {220, 18}

\bibitem[\protect\citeauthoryear{{Carrington}}{{Carrington}}{1859}]{Carrington_1859}
{Carrington} R.~C.,  1859, \mn@doi [\mnras] {10.1093/mnras/20.1.13}, \href
  {http://adsabs.harvard.edu/abs/1859MNRAS..20...13C} {20, 13}

\bibitem[\protect\citeauthoryear{{Cutri} \& {et al.}}{{Cutri} \& {et
  al.}}{2014}]{ALLWISE2014}
{Cutri} R.~M.,  {et al.} 2014, VizieR Online Data Catalog, \href
  {http://cdsads.u-strasbg.fr/abs/2014yCat.2328....0C} {2328}

\bibitem[\protect\citeauthoryear{{Davenport} et~al.,}{{Davenport}
  et~al.}{2014}]{Davenport14}
{Davenport} J. R.~A.,  et~al., 2014, \mn@doi [\apj]
  {10.1088/0004-637X/797/2/122}, \href
  {https://ui.adsabs.harvard.edu/\#abs/2014ApJ...797..122D} {797, 122}

\bibitem[\protect\citeauthoryear{{Dawson} et~al.,}{{Dawson}
  et~al.}{2013}]{Dawson13}
{Dawson} K.~S.,  et~al., 2013, \mn@doi [\aj] {10.1088/0004-6256/145/1/10},
  \href {https://ui.adsabs.harvard.edu/\#abs/2013AJ....145...10D} {145, 10}

\bibitem[\protect\citeauthoryear{{Dupuy} \& {Liu}}{{Dupuy} \&
  {Liu}}{2012}]{Dupuy12}
{Dupuy} T.~J.,  {Liu} M.~C.,  2012, \mn@doi [The Astrophysical Journal
  Supplement Series] {10.1088/0067-0049/201/2/19}, \href
  {https://ui.adsabs.harvard.edu/\#abs/2012ApJS..201...19D} {201, 19}

\bibitem[\protect\citeauthoryear{{Gaia Collaboration} et~al.,}{{Gaia
  Collaboration} et~al.}{2018}]{Gaia18}
{Gaia Collaboration} et~al., 2018, \mn@doi [\aap]
  {10.1051/0004-6361/201833051}, \href
  {https://ui.adsabs.harvard.edu/#abs/2018A&A...616A...1G} {616, A1}

\bibitem[\protect\citeauthoryear{{Gillen}, {Hillenbrand}, {David}, {Aigrain},
  {Rebull}, {Stauffer}, {Cody}  \& {Queloz}}{{Gillen} et~al.}{2017}]{Gillen17}
{Gillen} E.,  {Hillenbrand} L.~A.,  {David} T.~J.,  {Aigrain} S.,  {Rebull} L.,
   {Stauffer} J.,  {Cody} A.~M.,   {Queloz} D.,  2017, \mn@doi [\apj]
  {10.3847/1538-4357/aa84b3}, \href
  {https://ui.adsabs.harvard.edu/\#abs/2017ApJ...849...11G} {849, 11}

\bibitem[\protect\citeauthoryear{{Gizis}, {Burgasser}, {Berger}, {Williams},
  {Vrba}, {Cruz}  \& {Metchev}}{{Gizis} et~al.}{2013}]{Gizis13}
{Gizis} J.~E.,  {Burgasser} A.~J.,  {Berger} E.,  {Williams} P. K.~G.,  {Vrba}
  F.~J.,  {Cruz} K.~L.,   {Metchev} S.,  2013, \mn@doi [\apj]
  {10.1088/0004-637X/779/2/172}, \href
  {https://ui.adsabs.harvard.edu/\#abs/2013ApJ...779..172G} {779, 172}

\bibitem[\protect\citeauthoryear{{Gizis} et~al.,}{{Gizis}
  et~al.}{2015}]{Gizis_clouds}
{Gizis} J.~E.,  et~al., 2015, \mn@doi [\apj] {10.1088/0004-637X/813/2/104},
  \href {https://ui.adsabs.harvard.edu/\#abs/2015ApJ...813..104G} {813, 104}

\bibitem[\protect\citeauthoryear{{Gizis}, {Paudel}, {Schmidt}, {Williams}  \&
  {Burgasser}}{{Gizis} et~al.}{2017a}]{Gizis17}
{Gizis} J.~E.,  {Paudel} R.~R.,  {Schmidt} S.~J.,  {Williams} P. K.~G.,
  {Burgasser} A.~J.,  2017a, \mn@doi [\apj] {10.3847/1538-4357/aa6197}, \href
  {https://ui.adsabs.harvard.edu/\#abs/2017ApJ...838...22G} {838, 22}

\bibitem[\protect\citeauthoryear{{Gizis}, {Paudel}, {Mullan}, {Schmidt},
  {Burgasser}  \& {Williams}}{{Gizis} et~al.}{2017b}]{Gizis17b}
{Gizis} J.~E.,  {Paudel} R.~R.,  {Mullan} D.,  {Schmidt} S.~J.,  {Burgasser}
  A.~J.,   {Williams} P. K.~G.,  2017b, \mn@doi [\apj]
  {10.3847/1538-4357/aa7da0}, \href
  {https://ui.adsabs.harvard.edu/\#abs/2017ApJ...845...33G} {845, 33}

\bibitem[\protect\citeauthoryear{{Hawley} \& {Fisher}}{{Hawley} \&
  {Fisher}}{1992}]{Hawley1992}
{Hawley} S.~L.,  {Fisher} G.~H.,  1992, \mn@doi [\apjs] {10.1086/191640}, \href
  {http://adsabs.harvard.edu/abs/1992ApJS...78..565H} {78, 565}

\bibitem[\protect\citeauthoryear{{Hawley} et~al.,}{{Hawley}
  et~al.}{2003}]{Hawley04}
{Hawley} S.~L.,  et~al., 2003, \mn@doi [\apj] {10.1086/378351}, \href
  {https://ui.adsabs.harvard.edu/\#abs/2003ApJ...597..535H} {597, 535}

\bibitem[\protect\citeauthoryear{{Hawley}, {Davenport}, {Kowalski},
  {Wisniewski}, {Hebb}, {Deitrick}  \& {Hilton}}{{Hawley}
  et~al.}{2014}]{Hawley14}
{Hawley} S.~L.,  {Davenport} J. R.~A.,  {Kowalski} A.~F.,  {Wisniewski} J.~P.,
  {Hebb} L.,  {Deitrick} R.,   {Hilton} E.~J.,  2014, \mn@doi [\apj]
  {10.1088/0004-637X/797/2/121}, \href
  {https://ui.adsabs.harvard.edu/\#abs/2014ApJ...797..121H} {797, 121}

\bibitem[\protect\citeauthoryear{{Howell} et~al.,}{{Howell}
  et~al.}{2014}]{Howell14}
{Howell} S.~B.,  et~al., 2014, \mn@doi [Publications of the Astronomical
  Society of the Pacific] {10.1086/676406}, \href
  {https://ui.adsabs.harvard.edu/\#abs/2014PASP..126..398H} {126, 398}

\bibitem[\protect\citeauthoryear{{Jackman} et~al.,}{{Jackman}
  et~al.}{2018}]{Jackman18a}
{Jackman} J. A.~G.,  et~al., 2018, \mn@doi [\mnras] {10.1093/mnras/sty897},
  \href {https://ui.adsabs.harvard.edu/\#abs/2018MNRAS.477.4655J} {477, 4655}

\bibitem[\protect\citeauthoryear{{Kao}, {Hallinan}, {Pineda}, {Stevenson}  \&
  {Burgasser}}{{Kao} et~al.}{2018}]{Kao18}
{Kao} M.~M.,  {Hallinan} G.,  {Pineda} J.~S.,  {Stevenson} D.,   {Burgasser}
  A.,  2018, \mn@doi [The Astrophysical Journal Supplement Series]
  {10.3847/1538-4365/aac2d5}, \href
  {https://ui.adsabs.harvard.edu/\#abs/2018ApJS..237...25K} {237, 25}

\bibitem[\protect\citeauthoryear{{Lawrence} et~al.,}{{Lawrence}
  et~al.}{2007}]{Lawrence07}
{Lawrence} A.,  et~al., 2007, \mn@doi [\mnras]
  {10.1111/j.1365-2966.2007.12040.x}, \href
  {https://ui.adsabs.harvard.edu/\#abs/2007MNRAS.379.1599L} {379, 1599}

\bibitem[\protect\citeauthoryear{{Liebert}, {Kirkpatrick}, {Cruz}, {Reid},
  {Burgasser}, {Tinney}  \& {Gizis}}{{Liebert} et~al.}{2003}]{Liebert03}
{Liebert} J.,  {Kirkpatrick} J.~D.,  {Cruz} K.~L.,  {Reid} I.~N.,  {Burgasser}
  A.,  {Tinney} C.~G.,   {Gizis} J.~E.,  2003, \mn@doi [\aj] {10.1086/345514},
  \href {http://adsabs.harvard.edu/abs/2003AJ....125..343L} {125, 343}

\bibitem[\protect\citeauthoryear{{Miles-P{\'a}ez}, {Metchev}, {Heinze}  \&
  {Apai}}{{Miles-P{\'a}ez} et~al.}{2017}]{Miles17}
{Miles-P{\'a}ez} P.~A.,  {Metchev} S.~A.,  {Heinze} A.,   {Apai} D.,  2017,
  \mn@doi [\apj] {10.3847/1538-4357/aa6f11}, \href
  {https://ui.adsabs.harvard.edu/\#abs/2017ApJ...840...83M} {840, 83}

\bibitem[\protect\citeauthoryear{{Moon}, {Choe}, {Park}, {Wang}, {Gallagher},
  {Chae}, {Yun}  \& {Goode}}{{Moon} et~al.}{2002}]{Moon02}
{Moon} Y.-J.,  {Choe} G.~S.,  {Park} Y.~D.,  {Wang} H.,  {Gallagher} P.~T.,
  {Chae} J.,  {Yun} H.~S.,   {Goode} P.~R.,  2002, \mn@doi [\apj]
  {10.1086/340945}, \href {http://adsabs.harvard.edu/abs/2002ApJ...574..434M}
  {574, 434}

\bibitem[\protect\citeauthoryear{{Muirhead}, {Dressing}, {Mann}, {Rojas-Ayala},
  {L{\'e}pine}, {Paegert}, {De Lee}  \& {Oelkers}}{{Muirhead}
  et~al.}{2018}]{Muirhead18}
{Muirhead} P.~S.,  {Dressing} C.~D.,  {Mann} A.~W.,  {Rojas-Ayala} B.,
  {L{\'e}pine} S.,  {Paegert} M.,  {De Lee} N.,   {Oelkers} R.,  2018, \mn@doi
  [\aj] {10.3847/1538-3881/aab710}, \href
  {https://ui.adsabs.harvard.edu/\#abs/2018AJ....155..180M} {155, 180}

\bibitem[\protect\citeauthoryear{{Osten} et~al.,}{{Osten}
  et~al.}{2010}]{Osten10}
{Osten} R.~A.,  et~al., 2010, \mn@doi [\apj] {10.1088/0004-637X/721/1/785},
  \href {https://ui.adsabs.harvard.edu/\#abs/2010ApJ...721..785O} {721, 785}

\bibitem[\protect\citeauthoryear{{Paudel}, {Gizis}, {Mullan}, {Schmidt},
  {Burgasser}, {Williams}  \& {Berger}}{{Paudel} et~al.}{2018}]{Paudel18}
{Paudel} R.~R.,  {Gizis} J.~E.,  {Mullan} D.~J.,  {Schmidt} S.~J.,  {Burgasser}
  A.~J.,  {Williams} P. K.~G.,   {Berger} E.,  2018, \mn@doi [\apj]
  {10.3847/1538-4357/aab8fe}, \href
  {https://ui.adsabs.harvard.edu/\#abs/2018ApJ...858...55P} {858, 55}

\bibitem[\protect\citeauthoryear{{P{\'e}rez-Garrido}, {Lodieu}  \&
  {Rebolo}}{{P{\'e}rez-Garrido} et~al.}{2017}]{Perez17}
{P{\'e}rez-Garrido} A.,  {Lodieu} N.,   {Rebolo} R.,  2017, \mn@doi [\aap]
  {10.1051/0004-6361/201628778}, \href
  {https://ui.adsabs.harvard.edu/\#abs/2017A&A...599A..78P} {599, A78}

\bibitem[\protect\citeauthoryear{{Pineda}, {Hallinan}  \& {Kao}}{{Pineda}
  et~al.}{2017}]{Pineda17}
{Pineda} J.~S.,  {Hallinan} G.,   {Kao} M.~M.,  2017, \mn@doi [\apj]
  {10.3847/1538-4357/aa8596}, \href
  {https://ui.adsabs.harvard.edu/\#abs/2017ApJ...846...75P} {846, 75}

\bibitem[\protect\citeauthoryear{{Ricker} et~al.,}{{Ricker}
  et~al.}{2015}]{Ricker15}
{Ricker} G.~R.,  et~al., 2015, \mn@doi [Journal of Astronomical Telescopes,
  Instruments, and Systems] {10.1117/1.JATIS.1.1.014003}, \href
  {https://ui.adsabs.harvard.edu/\#abs/2015JATIS...1a4003R} {1, 014003}

\bibitem[\protect\citeauthoryear{{Schmidt}, {West}, {Hawley}  \&
  {Pineda}}{{Schmidt} et~al.}{2010}]{Schmidt10}
{Schmidt} S.~J.,  {West} A.~A.,  {Hawley} S.~L.,   {Pineda} J.~S.,  2010,
  \mn@doi [\aj] {10.1088/0004-6256/139/5/1808}, \href
  {https://ui.adsabs.harvard.edu/\#abs/2010AJ....139.1808S} {139, 1808}

\bibitem[\protect\citeauthoryear{{Schmidt}, {Hawley}, {West}, {Bochanski},
  {Davenport}, {Ge}  \& {Schneider}}{{Schmidt} et~al.}{2015}]{Schmidt15}
{Schmidt} S.~J.,  {Hawley} S.~L.,  {West} A.~A.,  {Bochanski} J.~J.,
  {Davenport} J. R.~A.,  {Ge} J.,   {Schneider} D.~P.,  2015, \mn@doi [\aj]
  {10.1088/0004-6256/149/5/158}, \href
  {https://ui.adsabs.harvard.edu/\#abs/2015AJ....149..158S} {149, 158}

\bibitem[\protect\citeauthoryear{{Schmidt} et~al.,}{{Schmidt}
  et~al.}{2016}]{Schmidt16}
{Schmidt} S.~J.,  et~al., 2016, \mn@doi [\apjl] {10.3847/2041-8205/828/2/L22},
  \href {http://adsabs.harvard.edu/abs/2016ApJ...828L..22S} {828, L22}

\bibitem[\protect\citeauthoryear{{Schmidt} et~al.,}{{Schmidt}
  et~al.}{2018}]{Schmidt18}
{Schmidt} S.~J.,  et~al., 2018, arXiv e-prints, \href
  {https://ui.adsabs.harvard.edu/\#abs/2018arXiv180904510S} {p.
  arXiv:1809.04510}

\bibitem[\protect\citeauthoryear{{Shibayama} et~al.,}{{Shibayama}
  et~al.}{2013}]{Shibayama13}
{Shibayama} T.,  et~al., 2013, \mn@doi [The Astrophysical Journal Supplement
  Series] {10.1088/0067-0049/209/1/5}, \href
  {https://ui.adsabs.harvard.edu/#abs/2013ApJS..209....5S} {209, 5}

\bibitem[\protect\citeauthoryear{{Skrutskie} et~al.,}{{Skrutskie}
  et~al.}{2006}]{2MASS_2006}
{Skrutskie} M.~F.,  et~al., 2006, \mn@doi [\aj] {10.1086/498708}, \href
  {http://adsabs.harvard.edu/abs/2006AJ....131.1163S} {131, 1163}

\bibitem[\protect\citeauthoryear{{Skrzypek}, {Warren}  \& {Faherty}}{{Skrzypek}
  et~al.}{2016}]{Skrzypek16}
{Skrzypek} N.,  {Warren} S.~J.,   {Faherty} J.~K.,  2016, \mn@doi [\aap]
  {10.1051/0004-6361/201527359}, \href
  {https://ui.adsabs.harvard.edu/\#abs/2016A&A...589A..49S} {589, A49}

\bibitem[\protect\citeauthoryear{{Stephens} et~al.,}{{Stephens}
  et~al.}{2009}]{Stephens09}
{Stephens} D.~C.,  et~al., 2009, \mn@doi [\apj] {10.1088/0004-637X/702/1/154},
  \href {http://adsabs.harvard.edu/abs/2009ApJ...702..154S} {702, 154}

\bibitem[\protect\citeauthoryear{Tsurutani, Gonzalez, Lakhina  \&
  Alex}{Tsurutani et~al.}{2003}]{Carrington_Energy}
Tsurutani B.~T.,  Gonzalez W.~D.,  Lakhina G.~S.,   Alex S.,  2003, \mn@doi
  [Journal of Geophysical Research: Space Physics] {10.1029/2002JA009504}, 108,
  n/a

\bibitem[\protect\citeauthoryear{{Wheatley} et~al.,}{{Wheatley}
  et~al.}{2018}]{Wheatley18}
{Wheatley} P.~J.,  et~al., 2018, \mn@doi [\mnras] {10.1093/mnras/stx2836},
  \href {https://ui.adsabs.harvard.edu/#abs/2018MNRAS.475.4476W} {475, 4476}

\bibitem[\protect\citeauthoryear{{Wright} et~al.,}{{Wright}
  et~al.}{2010}]{Wright10}
{Wright} E.~L.,  et~al., 2010, \mn@doi [\aj] {10.1088/0004-6256/140/6/1868},
  \href {https://ui.adsabs.harvard.edu/\#abs/2010AJ....140.1868W} {140, 1868}

\bibitem[\protect\citeauthoryear{{Yang}, {Liu}, {Qiao}, {Zhang}, {Gao}, {Cui}
  \& {Han}}{{Yang} et~al.}{2018}]{Yang18}
{Yang} H.,  {Liu} J.,  {Qiao} E.,  {Zhang} H.,  {Gao} Q.,  {Cui} K.,   {Han}
  H.,  2018, \mn@doi [\apj] {10.3847/1538-4357/aabd31}, \href
  {https://ui.adsabs.harvard.edu/\#abs/2018ApJ...859...87Y} {859, 87}

\bibitem[\protect\citeauthoryear{{York} et~al.,}{{York}
  et~al.}{2000}]{York2000}
{York} D.~G.,  et~al., 2000, \mn@doi [\aj] {10.1086/301513}, \href
  {https://ui.adsabs.harvard.edu/\#abs/2000AJ....120.1579Y} {120, 1579}

\makeatother
\end{thebibliography}
